\documentclass[preprint]{elsarticle}
\usepackage{graphicx,epsfig}
\usepackage{amsmath,amssymb}
\usepackage{lineno,hyperref}
\hypersetup{colorlinks=true}
\usepackage{graphicx}
\usepackage{wrapfig}
\usepackage[font=small]{caption}
\usepackage{subfigure}
\modulolinenumbers[2]
\biboptions{sort&compress}

\renewcommand{\vec}[1]{\ensuremath{\mathbf{#1}}} 
\newcommand{\gv}[1]{\ensuremath{\mbox{\boldmath$ #1 $}}} 
\newcommand{\dd}{\mathrm{d}} 
\newcommand{\ii}{\mathrm{i}} 
\newcommand{\eee}[1]{\mathrm{e}^{#1}} 
\newcommand{\pd}[2]{\frac{\partial #1}{\partial #2}} 
 
\renewcommand{\phi}{\varphi}

\newcommand{\bfGamma}{\pmb{\Gamma}}
\newcommand{\gtt}{g_{00}}
\newcommand{\grr}{g_{rr}}

\newcommand{\rs}{\mathrm{r_s}}

\journal{Physics Letters A}

\begin{document}

\begin{frontmatter}

\title{Anisotropic metamaterial as an analogue of a black hole}

\author{Isabel Fern\'{a}ndez-N\'{u}\~{n}ez}
\author{Oleg Bulashenko}

\address{Facultat de F\'{\i}sica, Universitat de Barcelona, Diagonal 645, 08028
Barcelona, Spain.}

\begin{abstract}
Propagation of light in a metamaterial medium which mimics curved spacetime and
acts like a black hole is studied.
We show that for a particular type of spacetimes and wave polarization, the time
dilation appears as dielectric permittivity, while the spatial curvature
manifests as magnetic permeability.
The optical analogue to the relativistic Hamiltonian which determines the ray
paths (null geodesics) in the anisotropic metamaterial is obtained.
By applying the formalism to the Schwarzschild metric, we compare the ray
paths with full-wave simulations in the equivalent optical medium.

\end{abstract}

\begin{keyword}
Metamaterials \sep Anisotropic optical materials \sep Analogue gravity \sep
Black hole \sep Schwarzschild spacetime 
\end{keyword}

\end{frontmatter}


\section{Introduction}
One of the hot topics of modern technology is to build artificial materials
whose permittivity and permeability can be properly engineered by incorporating
structural elements of subwavelength sizes. 
As a result, one can create materials (called {\em metamaterials})
with the desired electromagnetic response which offers new opportunities for
realizing such exotic phenomena as negative refraction, cloaking, super-lenses
for subwavelength imaging, microantennas, etc.
\cite{cai-shalaev-10,cui-smith-liu-10}.

Recently, it has also been recognized that metamaterials can be used to mimic
general-relativity phenomena
\cite{leonhardt0609, leonhardt10}.
The propagation of electromagnetic waves in curved spacetime is formally
equivalent to the propagation in flat spacetime in a certain inhomogeneous
anisotropic or bianisotropic medium
\cite{tamm25, skrotskii57, plebanski60, volkov70, felice71, mashhoon73,
schleich-scully84, thompson10}.
Based on this equivalence, different general relativity phenomena have been
discussed from the point of view of possible realization in metamaterials:
optical analogues of black holes
\cite{narimanov09, genov09, cheng10, yang12, yin13, sheng13},
Schwarzschild  spacetime \cite{thompson10,chen10},
de Sitter spacetime \cite{li10a, smolyaninov11a,mackay11},
cosmic strings \cite{mackay10, anderson10}, wormholes \cite{greenleaf07},
Hawking radiation \cite{schutzhold02}, the ``Big Bang'' and cosmological
inflation \cite{smolyaninov11b,smolyaninov12}, colliding gravitational waves
\cite{bini14}, among others.


The deflection of light waves in gradient-index optical materials mimicking
optical black holes was studied theoretically
\cite{narimanov09, genov09, sheng13} 
and experimentally \cite{cheng10, yang12, yin13, sheng13}.
These materials, called by authors ``omnidirectional electromagnetic
absorbers'', are characterized by an isotropic effective refractive index.
A real cosmological black hole (BH) can often be described by an anisotropic
spacetime, as, for example, the case of the Schwarzschild BH.
For that case, one should determine the permittivity and permeability tensors
instead of the refractive index in order to introduce the equivalent optical
medium \cite{thompson10, chen10}.
Chen et al.\ \cite{chen10} simulated the wave propagation outside the
Schwarzschild BH and observed in their numerical results the phenomenon of
``photon sphere'', which is an important feature of the BH system.
It would be interesting to go further and study the propagation of light waves
in optically anisotropic media which mimic cosmological BHs and compare the
results with ray paths obtained from the Hamiltonian method.

The aim of this letter is twofold.
First, we determine the constitutive relations of an inhomogeneous anisotropic
medium which is formally equivalent to the static spacetime metric obeying
rotational symmetries and can be applied, in principle, to the medium either in
isotropic or anisotropic form.
Second, by making use of the eikonal approximation to the wave equation, 
we obtain the expression for the optical Hamiltonian which we found to be
identical to the one obtained from general relativity for null geodesics, but
different from the optical Hamiltonian used in Refs.\
\cite{narimanov09, cheng10, jacob07, lee14}.
Then we apply the formalism to the Schwarzschild spacetime that is a solution to
the Einstein field equations in vacuum \cite{weinberg72}. We compare the wave
propagation with the ray dynamics outside the BH in the effective medium and
obtain a very good correspondence. As an interesting feature we find that light
does not propagate in the direction of the wave normal, there is an angle
between the wave velocity and the ray velocity. The obtained results are
discussed from the point of view of metamaterial implementation.

\section{General relativity in a metamaterial medium}\label{sec:metam}

\subsection{Medium parameters}

Long time ago, Tamm pointed out the parallels between anisotropic crystals and
curved spacetimes  \cite{tamm25}.
Later studies showed \cite{skrotskii57, plebanski60, volkov70, felice71,
mashhoon73, schleich-scully84}
that the propagation of light in empty curved space distorted by a gravitational
field is formally equivalent to light propagation in flat space filled with an
inhomogeneous anisotropic medium.

Indeed, consider a spacetime background with a general metric 
\footnote{From now on we follow the standard notations for covariant
(subindices) and contravariant (superindices) quantities.}
\begin{equation}
\dd s^2 = \gtt\,\dd t^2 + 2g_{0i}\,\dd t \,\dd x^i + g_{ij} \,\dd x^i \,\dd x^j,
\label{eq:gen-metric}
\end{equation}
where $i,j=1,2,3$ run over arbitrary spatial coordinates.
Then, it can be shown \cite{schleich-scully84} that the covariant Maxwell's
equations written in curved coordinates can be transformed into their standard
form for flat space but in the presence of an effective medium.
The constitutive relations of the equivalent medium have been found in the form
\cite{plebanski60}: 
\begin{equation}
D^i=\varepsilon^{ij}E_j-(\bfGamma \times \mathbf{H})^i, \quad
B^i=\mu^{ij}H_j+(\bfGamma \times \mathbf{E})^i,
\label{DBEH}
\end{equation}
which connect the fields $\mathbf{D}$, $\mathbf{B}$, $\mathbf{E}$ and
$\mathbf{H}$ via nontrivial permittivity and permeability tensors
\begin{equation}
\varepsilon^{ij}=\mu^{ij}=-\frac{\sqrt{-g}}{g_{00}}g^{ij}
\label{eq:perm}
\end{equation}
and a vector $\mathbf{\Gamma}$ given by
\begin{equation}
\label{eq:gam}
\Gamma_i = -  \frac{g_{0i}}{g_{00}}.
\end{equation}
Here, $g^{ij}$ is the inverse of $g_{ij}$ and $g$ is the determinant of the
full spacetime metric $g_{\mu\nu}$, with $\mu,\nu=0,1,2,3$.
Note that the information about the gravitational field is essentially embedded
in the material properties of the effective medium: the tensors
$\varepsilon^{ij}$, $\mu^{ij}$ which are symmetric and should be equal, and the
vector $\mathbf{\Gamma}$ which couples the electric and magnetic fields.
The invention of metamaterials during the last decade \cite{cai-shalaev-10,
cui-smith-liu-10} opened up the possibility to design electromagnetic media
corresponding to different spacetimes \cite{leonhardt0609, leonhardt10, chen10, 
li10a, smolyaninov11a, mackay11, mackay10, anderson10, greenleaf07,
schutzhold02, smolyaninov11b, smolyaninov12, bini14}.

In this letter, we consider a static spacetime metric associated with a
spherically symmetric cosmological BH.
Due to time-reversal symmetry, $g_{0i}=0$ and the coupling between the electric
and magnetic fields vanishes, $\mathbf{\Gamma}=0$.
The metric \eqref{eq:gen-metric} in $(t,r,\theta,\phi)$ coordinates can then be
written in a generic form as \cite{weinberg72}
\begin{equation}
\dd s^2 = \gtt(r) \, \dd t^2 + \grr(r)\, \left\{ \dd r^2  + f(r) \, [ r^2\,
\dd\theta^2 + r^2 \sin^2 \theta \, \dd\phi^2 ] \right\},
\label{eq:stat-spher-metric}
\end{equation}
where $f$ is the ``anisotropic factor''.
Note that the metric (\ref{eq:stat-spher-metric}) obeys the rotational
symmetries in the three-dimensional $(r,\theta,\phi)$ space.

Then, we have to project the metric \eqref{eq:stat-spher-metric} into a flat
background to obtain the medium parameters in the Cartesian coordinate system.
To do that, we apply a coordinate transformation and, from Eq.\ \eqref{eq:perm},
we get the permittivity and permeability tensors in the form:
\begin{equation}
\varepsilon^{ij} = \mu^{ij}= \sqrt{- \frac{\grr}{\gtt} }
\left[ \delta^{ij} - (1-f) \frac{x^i x^j}{r^2} \right],
\label{eq:perm-gen}
\end{equation}
where $\delta^{ij}$ is the Kronecker delta, $r=\sqrt{x^2+y^2+z^2}$, and we
denoted the Cartesian coordinates $(x^1, x^2, x^3) \equiv (x,y,z)$.
It is seen that whenever $f \neq 1$, the permittivity and permeability tensors
contain the off-diagonal elements and the equivalent medium is essentially
anisotropic.
Only in the case of $f=1$, the off-diagonal elements vanish and the medium
becomes completely isotropic with all the diagonal elements equal to the
refractive index: $n(r)=\sqrt{-\grr/\gtt}$.
Note that in general relativity the spacetime with $f=1$ in Eq.\
\eqref{eq:stat-spher-metric} is said to be conformal to flat space.
Every static spherically symmetric spacetime with $f \neq 1$ can be converted to
conformally flat form by an appropriate transformation of the radial coordinate:
$r \to \rho$. The new radial coordinate is obtained by
\begin{equation}
\rho=r \exp \left\{ \int_r^\infty \left[1- \frac{1}{\sqrt{f(r')} } \right]
\frac{\dd r'}{r'} \right\},
\label{eq:r-rho}
\end{equation}
where the isotropic boundary condition at infinity, $f(\infty)=1$, is taken into
account. The line element in the $(t,\rho,\theta,\phi)$ {\em isotropic}
coordinates takes the conformally flat form:
\begin{equation}
\dd s^2=\gtt[r(\rho)] \, \dd t^2+\Lambda(\rho) \, (\dd\rho^2 +
\rho^2\sin^2 \theta \, \dd\phi^2  + \rho^2\dd\theta^2),
\end{equation}
where the time dilation term $\gtt$ and the conformal factor 
$\Lambda=\grr f r^2/\rho^2$ are calculated by means of the function $r(\rho)$
which should be obtained by inverting \eqref{eq:r-rho}.
Thus, the permittivity and permeability tensors are simply reduced to the
isotropic refractive index:
\begin{equation}
\varepsilon^{ij} = \mu^{ij} =  \delta^{ij}
\sqrt{ - \frac{\Lambda}{\gtt} } \equiv \delta^{ij} \,n(\rho).
\label{eq:n-gen}
\end{equation}
The equivalent medium determined by Eq.\ \eqref{eq:n-gen} is still inhomogeneous
since the refractive index varies with the radial coordinate, but the light
velocity in the medium becomes isotropic, a property that is much simpler to
implement in metamaterial design.

In what follows, we will compare the results for light propagation in two
different equivalent media -- isotropic and anisotropic -- both corresponding
to the same spacetime metric in order to see the physical differences.

\subsection{Electromagnetic fields. TE and TM polarizations}
\label{TE-TM}

The results from the previous section indicate that an electromagnetic field can
be thought of as propagating in flat background but in the presence of a medium
whose properties are constructed from a curved spacetime.
The fields, for the static case we consider, are related by:
\begin{equation}
D^i=\varepsilon^{ij}E_j, \quad B^i=\mu^{ij}H_j.
\label{DBEH-stat}
\end{equation}
Due to the anisotropy of the medium for the metric in nonconformally flat form,
the electric displacement field $\mathbf{D}$ is not in the direction of
$\mathbf{E}$, and the magnetic induction field $\mathbf{B}$ is not in the
direction of $\mathbf{H}$.
To simplify the treatment of the problem, we consider the propagation of light
in the equatorial plane, $z=0$. In such a case, one of the anisotropies --
electric or magnetic -- can be eliminated.

Indeed, consider the TE polarization for an electromagnetic wave for which
$\vec{E}$ is perpendicular to the $x$-$y$ plane. Equation \eqref{eq:perm-gen}
for $z=0$ leads to $\varepsilon^{xz}=\varepsilon^{yz}=0$, hence the directions
of $\vec{D}$ and $\vec{E}$ coincide. This means that $\varepsilon^{zz}$ is the
only relevant matrix element which connects the nonzero electric components of
the field: $D^z=\varepsilon^{zz}E_z$, and the electric anisotropy of the medium
is irrelevant.
As for the magnetic components, we obtain $\mu^{xz}=\mu^{yz}=0$, $B^z=H^z=0$, 
and $B^x=\mu^{xx} H_x + \mu^{xy} H_y$ and $B^y=\mu^{xy} H_x + \mu^{yy} H_y$.
Hence, $\vec{B}$ and $\vec{H}$ are confined to the $x$-$y$ plane, but their
directions do not coincide.

\begin{figure}[h!]
     \centering
  \subfigure[\label{fig:angle-te}Magnetic anisotropy.]
  {\includegraphics[width=0.3\textwidth]{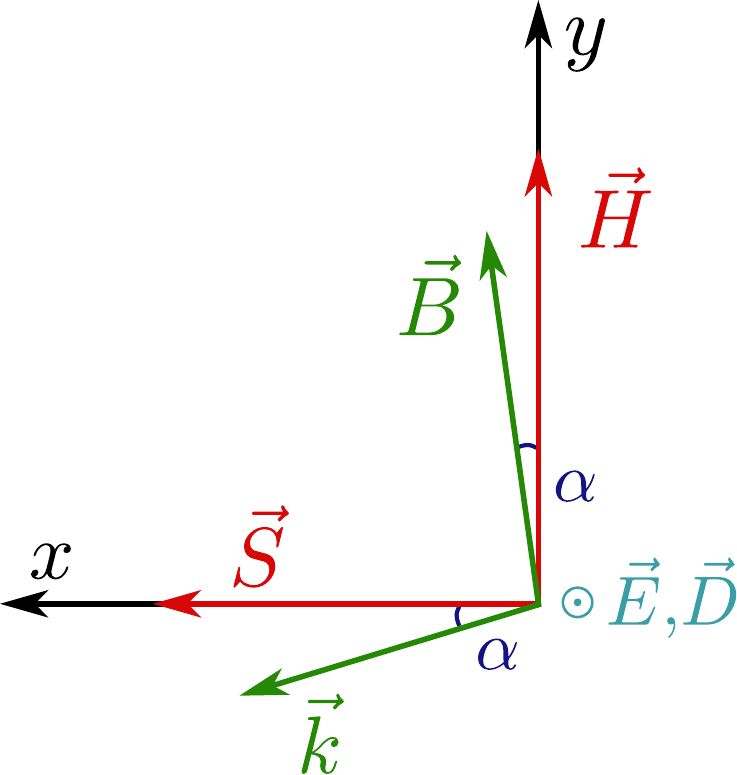}}
     \qquad
  \subfigure[\label{fig:angle-tm}Electric anisotropy.]
  {\includegraphics[width=0.3\textwidth]{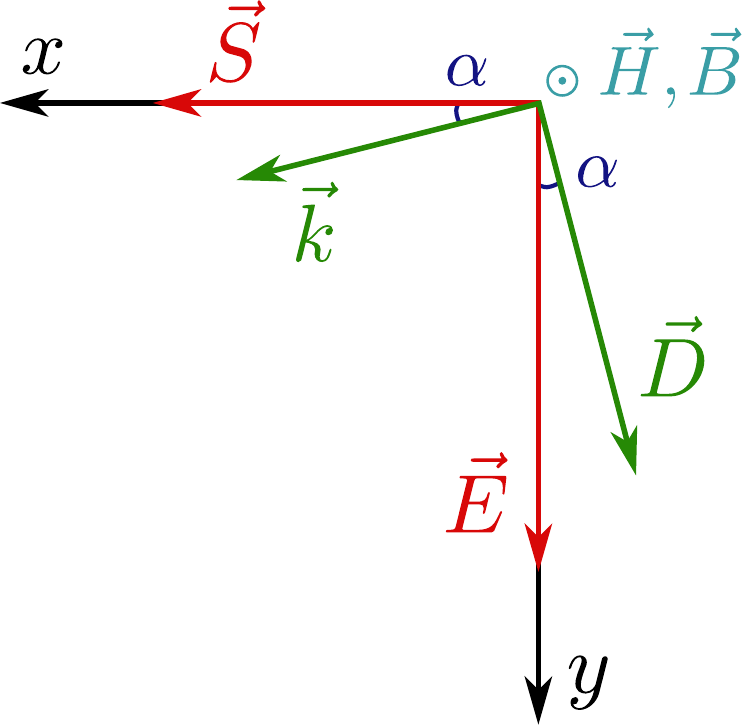}}
\caption{Electromagnetic fields $\vec{H}$, $\vec{B}$, $\vec{E}$, $\vec{D}$,
wavevector $\vec{k}$ and Poynting vector $\vec{S}$ for:
(a) TE polarization in a magnetically anisotropic medium, and (b) TM
polarization in an electrically anisotropic medium.}
\label{fig:angle}
\end{figure}

Similarly, one can consider the TM polarization for which $\vec{H}$ is
perpendicular to the $x$-$y$ plane. In such a case, one gets  $B^z=\mu^{zz}
H_z$, $D^x=\varepsilon^{xx} E_x + \varepsilon^{xy} E_y$, and
$D^y=\varepsilon^{xy} E_x + \varepsilon^{yy} E_y$.
This means that the directions of $\vec{B}$ and $\vec{H}$ coincide, and the
magnetic anisotropy is irrelevant.
It should be noted that, since $\varepsilon^{ij} = \mu^{ij}$ and they are both
symmetrical, the TE and TM polarizations are completely equivalent.
These two cases are summarized in Fig.\ref{fig:angle}.

The medium anisotropy has very important physical consequences on the
propagation of the electromagnetic waves. If we look at the monochromatic plane
wave with wavevector $\vec{k}$, then $\vec{D}$, $\vec{B}$ and $\vec{k}$
form an orthogonal vector triplet. On the other hand, $\vec{E}$, $\vec{H}$ and 
the Poynting vector $\vec{S}$ form another orthogonal vector triplet. Therefore,
the energy does not propagate in the direction of the wave normal
\cite{born-wolf-03}
and the ray velocity and the wave velocity are not equal, unlike what happens
for an isotropic medium. It is seen that the angle between $\vec{S}$ and
$\vec{k}$ (denoted by $\alpha$ in Fig.\ref{fig:angle}) is equal to that between
$\vec{B}$ and $\vec{H}$ in the TE polarization or between $\vec{D}$ and
$\vec{E}$ in the TM polarization. For the angle $\alpha$ we find:
\begin{equation}
\tan\alpha = - \frac{\mu^{xy}}{\mu^{yy}} = -
\frac{\varepsilon^{xy}}{\varepsilon^{yy}} = \frac{(1-f)xy}{x^2+f y^2}
\label{tan}
\end{equation}
Only when $f=1$, the medium becomes isotropic, $\alpha$ vanishes and the ray and
wave velocities become equal.

\subsection{Schwarzschild spacetime}

Equations \eqref{eq:perm-gen} and \eqref{eq:n-gen} can now be applied to any
static spherically symmetric spacetime.
As an example, we consider the Schwarzschild metric that is a solution to the
Einstein field equations in vacuum \cite{weinberg72}.
This metric in spherical coordinates reads:
\begin{equation}
\dd s^2 = - \left(1-\frac{\rs}{r}\right) \dd t^2 + \frac{\dd
r^2}{1-\frac{\rs}{r}} + r^2(\sin^2\theta \, \dd \phi^2  + \dd\theta^2).
\label{eq:sch-metric}
\end{equation}
Here, the speed of light $c = 1$, the Schwarzschild radius $r_s = 2GM/c^2$, $G$
is the universal gravitational constant and $M$ is the mass of the source of
gravitation.
Comparing with Eq.\ \eqref{eq:stat-spher-metric}, one can easily identify:
$\gtt = -(1-r_s/r)$, $\grr=-1/\gtt$, and $1-f=r_s/r$.
Since the Schwarzschild metric \eqref{eq:sch-metric} is not in a conformally
flat form, one obtains the anisotropic equivalent medium tensors
\begin{equation}
\varepsilon^{ij} = \mu^{ij}= \frac{1}{1-\frac{\rs}{r}}
\left( \delta^{ij} - \frac{x^i x^j}{r^3}\rs \right),
\label{eq:perm-sch}
\end{equation}
and from Eq.\ \eqref{tan} we find the angle between the wave and ray velocities:
\begin{equation}
\tan\alpha=\frac{xy\rs}{r^3-y^2\rs}
\label{tan-schw}
\end{equation}
Equation \eqref{eq:perm-sch} is a compact version of the medium tensors which
can also be found as matrices in Refs.\ \cite{chen10,mackay11}.

The Schwarzschild metric can be written in a conformally flat form by means of
the well known substitution (see, e.g. \cite{weinberg72})
\begin{equation}
r=\rho\left(1+\frac{\rs}{4\rho}\right)^2.
\label{eq:iso-r}
\end{equation}
The inverted formula for the isotropic radial coordinate $\rho$ reads:
\begin{equation}
\rho = \frac{1}{2} \left( r-\frac{1}{2} r_s + \sqrt{r^2 - r_s r} \right)
\label{eq:iso-rho}
\end{equation}
which can also be obtained from Eq.\ \eqref{eq:r-rho} by substituting
$f=1-\rs/r$.
In isotropic coordinates, the Schwarzschild metric becomes \cite{weinberg72}
\begin{equation}
\dd s^2 = - \left(\frac{1-\frac{\rs}{4\rho}}{1+\frac{\rs}{4\rho}}\right)^2\dd
t^2 + \left(1+\frac{\rs}{4\rho}\right)^4(\dd\rho^2 + \rho^2\sin^2\theta \,
\dd\phi^2  + \rho^2\dd\theta^2) .
\label{eq:iso-metric}
\end{equation}
Finally, the refractive index of the equivalent medium is found from
Eq.\ \eqref{eq:n-gen} as (see also Ref.\ \cite{felice71})
\begin{equation}
n(\rho)=\frac{\left(1+\frac{\rs}{4\rho}\right)^3}{1-\frac{\rs}{4\rho}},
\label{eq:n-iso}
\end{equation}
The formulas given in this section for the Schwarzschild spacetime will be used
in Sec.\ \ref{sec:results} for numerical simulation.

\section{Maxwell's equations in anisotropic medium}

The starting point of our analysis of wave propagation in the equivalent medium
is the Maxwell equations written in a source-free form
\begin{equation}
\gv{\nabla}\cdot\vec{B}=0, \quad
\gv{\nabla}\cdot\vec{D}=0, \quad
\gv{\nabla}\times\vec{E}=-\pd{\vec{B}}{t}, \quad
\gv{\nabla}\times\vec{H}=\pd{\vec{D}}{t}.
\label{eq:maxwell}
\end{equation}
These equations should be supplemented by the constitutive relations 
\eqref{DBEH-stat} with the medium tensors \eqref{eq:perm-gen} expressed through
the spacetime metric coefficients.
If the operating frequency bandwidth is sufficiently narrow, the dispersion can
be neglected and one can consider a monochromatic wave with frequency $\omega$.
Equations \eqref{eq:maxwell} can then be reduced \cite{landau-v8} to the wave
equation for a time-harmonic electric field vector $\vec{E}_{\omega}$
\begin{equation}
\gv{\nabla} \times
\left[ \underline{\underline{\mu}}^{-1} ( \gv{\nabla} \times \vec{E}_{\omega} )
\right] = \omega^2 \underline{\underline{\varepsilon}} \,\vec{E}_{\omega}.
\label{eq:wave}
\end{equation}
The medium is generally anisotropic, hence,
$\underline{\underline{\varepsilon}}$ and $\underline{\underline{\mu}}^{-1}$
denote the permittivity and inverse permeability tensors, respectively.

As it was discussed in Sec.\ \ref{TE-TM}, the electromagnetic propagation in the
equatorial plane, $z=0$, can be described in terms of either TE or TM waves. For
the TE wave, $\vec{E}_{\omega}(x,y,z)= E(x,y)\eee{-\ii k_z z}\vec{\hat{z}}$, we
obtain the equation for the $z$-component of the electric field $E$ in the
two-dimensional ($x,y$) space
\begin{equation}
\pd{}{x}\left(\mu_{xy}\pd{E}{y}-\mu_{yy}\pd{E}{x}\right)-\pd{}{y}\left(\mu_{xx}
\pd{E}{y}-\mu_{xy}\pd{E}{x}\right)=\omega^2\varepsilon^{zz}E,
\label{eq:te}
\end{equation}
where $\mu_{xx}, \mu_{xy}, \mu_{yy}$ are the corresponding components of the
tensor $\underline{\underline{\mu}}^{-1}$ and $\varepsilon^{zz}$ corresponds to
$\underline{\underline{\varepsilon}}$.
Since only one component of $\underline{\underline{\varepsilon}}$ has entered
into the final equation \eqref{eq:te}, the permittivity can be chosen isotropic
with all its diagonal elements equal to $\varepsilon^{zz}$ and the off-diagonal
terms equal to zero. In this case, the equivalent medium will be magnetically
anisotropic [see Fig. \ref{fig:angle-te}].

The wave equation for the TM case can also be easily obtained from the above
formulas by means of the substitutions:
$\varepsilon^{ij} \rightleftarrows \mu^{ij}$, $E\to H$.
In such a case, the medium is electrically anisotropic [Fig.
\ref{fig:angle-tm}], and one should solve the equation for the magnetic field
$H(x,y)$.

We would like to point out that we do not use the plane wave approximation to
treat the problem, instead, we will solve the wave equation  \eqref{eq:te}
numerically for a Gaussian beam with the appropriate boundary conditions. This
situation, as we believe, is more realistic from the experimental point of view.

\section{Hamiltonian formulation}

In the limit of geometrical optics, for which the wavelength is much smaller
than the scale of variation of the medium parameters, electromagnetic waves
should follow ray trajectories (null geodesics of the background spacetime).
To test the validity of our scheme, it would be interesting to compare the
full-wave numerical calculation of the propagation of light with its geometrical
optics limit, for which analytical solutions exist.
To accomplish this task, we have to define the Hamiltonian.

Let us start with the wave equation \eqref{eq:wave}, which we apply again to
the equatorial plane, $z=0$, and consider the TE polarization for the wave
propagation. 
Given that the medium tensors are diagonal in the spherical coordinate system,
it is advantageous to rewrite Eq.\ \eqref{eq:te} in polar coordinates:
$E(x,y)\to E(r,\phi)$. One gets, 
\begin{equation}
\frac{1}{r} \pd{}{r} \left( \frac{r}{\mu_{\phi}} \pd{E}{r} \right)
+ \frac{1}{r^2} \pd{}{\phi} \left( \frac{1}{\mu_r} \pd{E}{\phi} \right)
+ \omega^2 \varepsilon_z E=0.
\label{eq:te-r}
\end{equation}
where we denote $\mu_r\equiv\mu^{rr}$, $\mu_\phi\equiv\mu^{\phi\phi}$, and
$\varepsilon_z\equiv\varepsilon^{zz}$, which are the medium parameters in polar
coordinates.

In the eikonal approximation we seek the solution in the form:
$E(r,\phi) =E_0 \exp [ iS(r,\phi) ]$, where $E_0$ is the slowly varying
amplitude and $S(r,\phi)$ is the fast varying eikonal \cite{landau-v2}.
Substituting into Eq.\ \eqref{eq:te-r} and neglecting the small terms
corresponding to the derivatives over slowly varying functions, we obtain:
\begin{equation}
\frac{1}{\mu_{\phi}} \left( \pd{S}{r} \right)^2
+ \frac{1}{\mu_r r^2 } \left( \pd{S}{\phi} \right)^2
- \varepsilon_z \,\omega^2 =0,
\label{eq:eik-em}
\end{equation}
which can be rewritten in terms of the spacetime metric coefficients as
\begin{equation}
\frac{1}{\grr} \left[ \left( \pd{S}{r} \right)^2
+ \frac{1}{f r^2} \left( \pd{S}{\phi} \right)^2 \right]
+ \frac{1}{\gtt} \,\omega^2 =0.
\label{eq:eik-g}
\end{equation}
But this is nothing else than the eikonal equation for a light ray in a
gravitational field \cite{landau-v2}
\begin{equation}
g^{\alpha\beta}\pd{S}{x^{\alpha}} \pd{S}{x^{\beta}}=0
\label{eq:eik}
\end{equation}
with the indices $\alpha,\beta$ running over space and time.
The Hamiltonian representation of Eq.\ \eqref{eq:eik} is obtained by rewriting
it in terms of the variables $p_{\alpha}=\partial S/\partial x^{\alpha}$, where
$p_{\alpha}$ is the 4-momentum. One gets
\begin{equation}
\mathcal{H} = g^{\alpha\beta}p_\alpha p_\beta = p^\beta p_\beta = 0.
\label{H_0}
\end{equation}
Thus, with the definitions
$p_r=\partial S/\partial r$, $p_{\phi}=\partial S/\partial \phi$, and
$p_t=\omega$, just plugging them into Eq.\ \eqref{eq:eik-g}, we obtain
\begin{equation}
\mathcal{H}=\frac{1}{\gtt} \, p_t^2
+ \frac{1}{\grr} \left( p_r^2 + \frac{1}{f r^2} \, p_\phi^2 \right).
\label{Ham}
\end{equation}
The optic Hamiltonian $\mathcal{H}$ given by Eq.\ \eqref{Ham} describes the
ray trajectories in the equivalent anisotropic medium and is equivalent to the
one used to determine the null geodesics in the gravitational field.
It takes into account the time dilation by means of the term $\gtt$, the
spatial curvature by the metric coefficient $\grr$ and the medium anisotropy
through the function $f$. The latter should result in the deviation of the
momentum vector $\vec{p}$ from the ray propagation direction.

Having found the Hamiltonian $\mathcal{H}$, the ray dynamics can easily be
calculated from Hamilton's equations of motion:
\begin{equation}
\dot{q}^{\alpha} = \pd{\mathcal{H}}{p_{\alpha}}, \quad
\dot{p}_{\alpha} = - \pd{\mathcal{H}}{q^{\alpha}},
\label{Ham-eqs}
\end{equation}
where the canonical coordinates in our case are:
$q^{\alpha} = (t, r, \phi)$, $p_{\alpha} = (p_t, p_r, p_\phi)$
and the ``dot'' denotes the derivative over the affine parameter which varies
along the trajectory.
Note that in the Hamiltonian \eqref{Ham}, the coefficients $\gtt$, $\grr$
and $f$ are functions of only the radial coordinate. Therefore, there exist two
integrals of motion, $\dot{p}_t=0$ and $\dot{p}_{\phi}=0$, which just state the
conservation of energy and angular momentum for the static spherically
symmetric system. In principle, these two constants of motion can be introduced;
but in fact, since photons are massless, the ray trajectories are determined by
only one constant -- the impact parameter $b$ which is the ratio between them
\cite{chandra}.

We can now determine the ray trajectories for a particular spacetime metric by
means of Eqs.\ \eqref{Ham-eqs} with the appropriate boundary conditions.
To be specific, we consider the Schwarzschild metric \eqref{eq:sch-metric} for
which the Hamiltonian can be readily obtained by means of Eq.\ \eqref{Ham}
\begin{equation}
\mathcal{H}=-\frac{1}{1-\frac{\rs}{r}} \, p_t^2 + \left(1-\frac{\rs}{r}
\right)p_r^2+\frac{1}{r^2}\, p_\phi^2.
\label{eq:h-schw}
\end{equation}
On the other hand, in isotropic coordinates (see Eq.\ \eqref{eq:iso-metric}), 
the Schwarzschild Hamiltonian becomes 
\begin{equation}
\mathcal{H} = - \left( \frac{1+\frac{\rs}{4\rho}}{1-\frac{\rs}{4\rho}} \right)^2
p_t^2 + \frac{1}{ \left( 1 + \frac{\rs}{4\rho} \right)^4}
\left( p_\rho^2 + \frac{1}{\rho^2} \, p_\phi^2 \right).
\label{eq:h-iso}
\end{equation}
For both cases, isotropic and anisotropic, we will compare different kinds of
trajectories with the full-wave numerical simulations in
Sec.\ \ref{sec:results}.
Note that Eq.\ \eqref{H_0}, when applied to the optical medium, gives the
generalized dispersion relation \cite{leonhardt10} also known as the Fresnel
equation \cite{born-wolf-03, landau-v8}. The discussion of this relation for the
Schwarzschild equivalent medium can be found, e.g., in Ref.\ \cite{thompson10}.

It would be appropriate to make some important remarks here.

\begin{enumerate}
 \item[(i)]
By comparing the terms in Eqs.\ \eqref{eq:eik-em} and \eqref{eq:eik-g}, one
can see that for the TE wave, the time dilation term $\gtt$ is determined by
the dielectric permittivity of the equivalent medium, while the spatial metric
components $\grr$ and $f$ depend on the magnetic permeability.
In other words, the time and space metric coefficients correspond to different
properties of the medium: either electric or magnetic.
Similarly, if we consider the TM wave, one should replace: $\varepsilon^{ij}
\rightleftarrows \mu^{ij}$; $E\to H$ (see Sec.\ \ref{TE-TM}), and the
correspondence will be just the opposite: $\gtt$ is related to $\mu_z$, while
$\grr$ and $f$ correspond to $\varepsilon_r$ and $\varepsilon_{\phi}$.
 \item[(ii)]
Equation \eqref{eq:eik-em} can be divided by $\varepsilon_z$ and the medium
parameters can be redefined according to:
$\tilde{\mu}_{\phi}=\mu_{\phi}\varepsilon_z$, 
$\tilde{\mu}_r=\mu_r \varepsilon_z$ and $\tilde{\varepsilon}=1$.
It is easy to check that the new medium gives exactly the same ray
trajectories, although it has the permittivity $\tilde{\varepsilon}$ of free
space and the inhomogeneous permeability $\tilde{\mu}$.
Actually, the Hamiltonian $\mathcal{H}$ may be multiplied by any arbitrary
function of the coordinates (by $\gtt$, in this case) without changing the ray
path obtained from the equations of motion since only the parametrization of
the affine parameter changes.
This means that in the geometrical optics limit, one can introduce a set of
equivalent media, all giving the same light ray paths.
But the wave equation \eqref{eq:te-r} (or in general Eq.\ \eqref{eq:te}) cannot
be renormalized in this way, and therefore the interference patterns for those
analogous media will be different.

\end{enumerate}

\section{Comparison of wave propagation with ray dynamics}
\label{sec:results}

In this section we present the results obtained numerically for the propagation
of light waves in an anisotropic medium which mimics a static spherically
symmetric cosmological spacetime.
As an example, we consider the Schwarzschild spacetime with the goal to
compare the results for two sets of coordinates: spherical and isotropic.
On the other hand, to validate our theoretical approach, we wish to compare the
full-wave numerical simulation with the ray trajectories obtained within the
Hamiltonian framework.

\begin{figure}
 \centering
\subfigure{\includegraphics[width=0.85\textwidth]{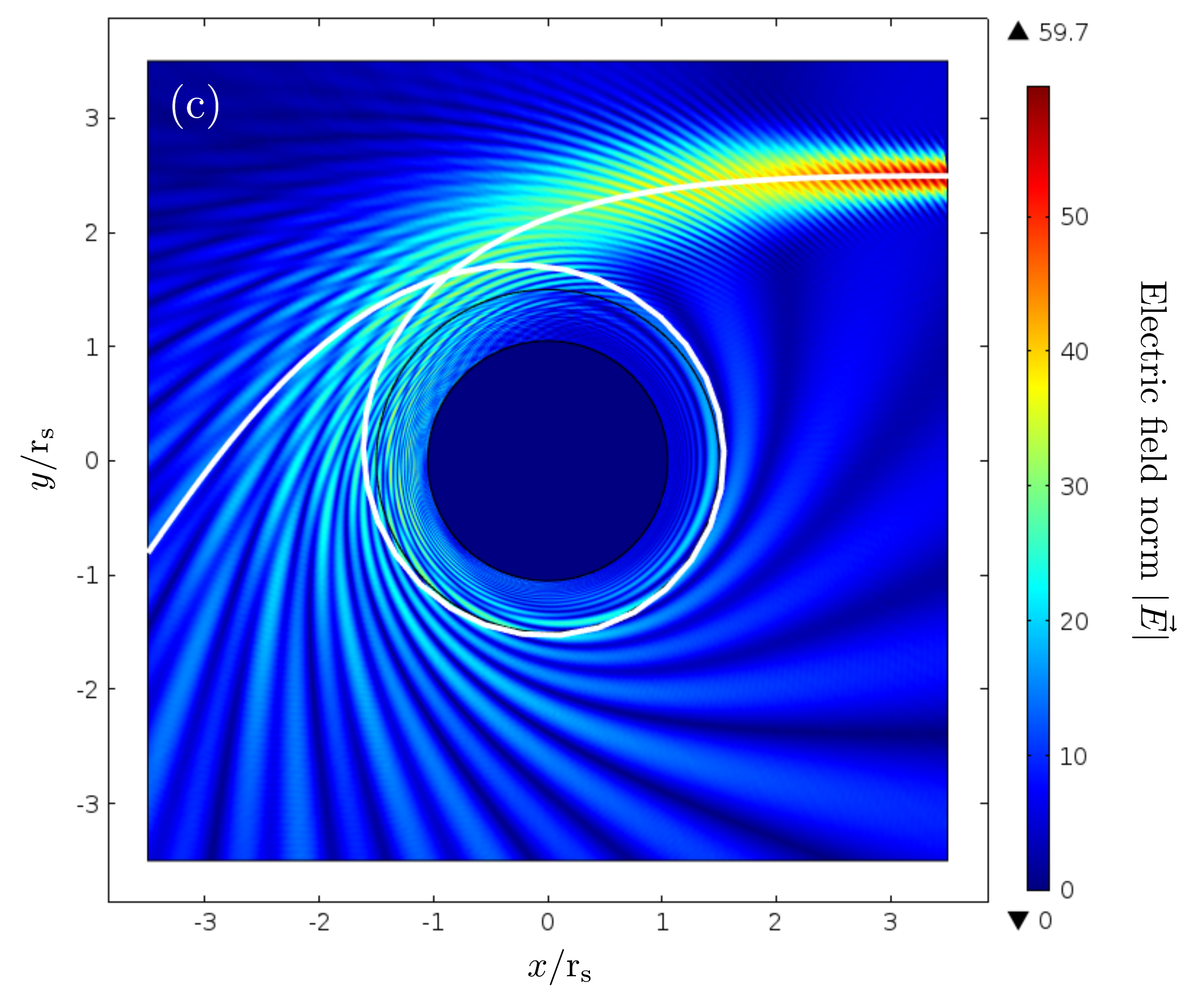}}
\subfigure{\includegraphics[width=0.85\textwidth]{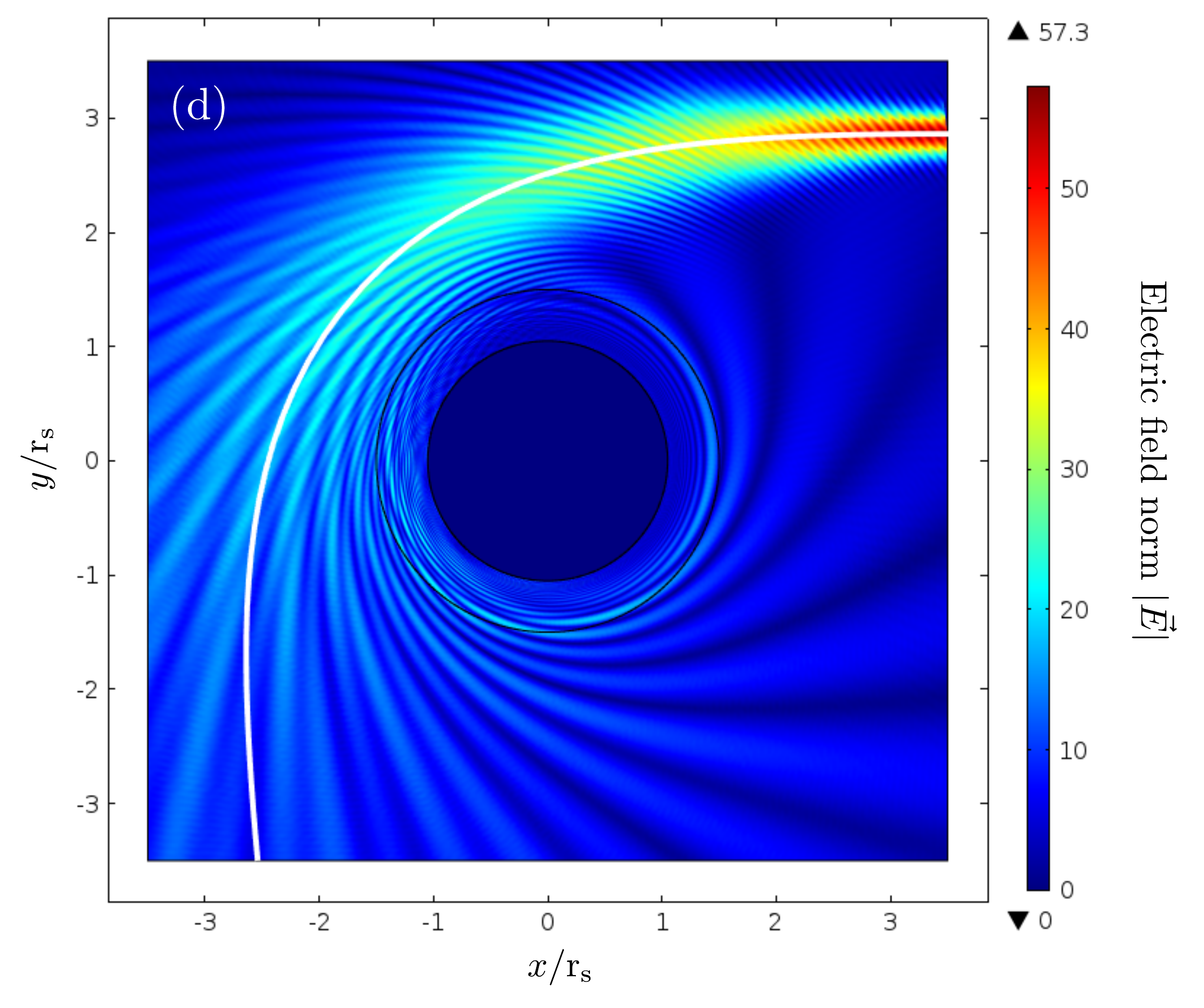}}
\caption{
TE Gaussian beam of wavelength $\lambda=0.15\rs$ compared with ray path 
(superimposed in white) in metamaterial mimicking the Schwarzschild
anisotropic spacetime for different impact parameters:
(a) $b=\rs$, (b) $b=2\rs$, (c) $b\approx b_{\rm c}$ (slightly above), and (d)
$b=3\rs$.
The event horizon $\rs$ and the photon sphere of radius $3\rs/2$
are depicted by black circles.}
\label{fig:schw}
\end{figure}

\begin{figure}
 \centering
\subfigure{\includegraphics[width=0.85\textwidth]{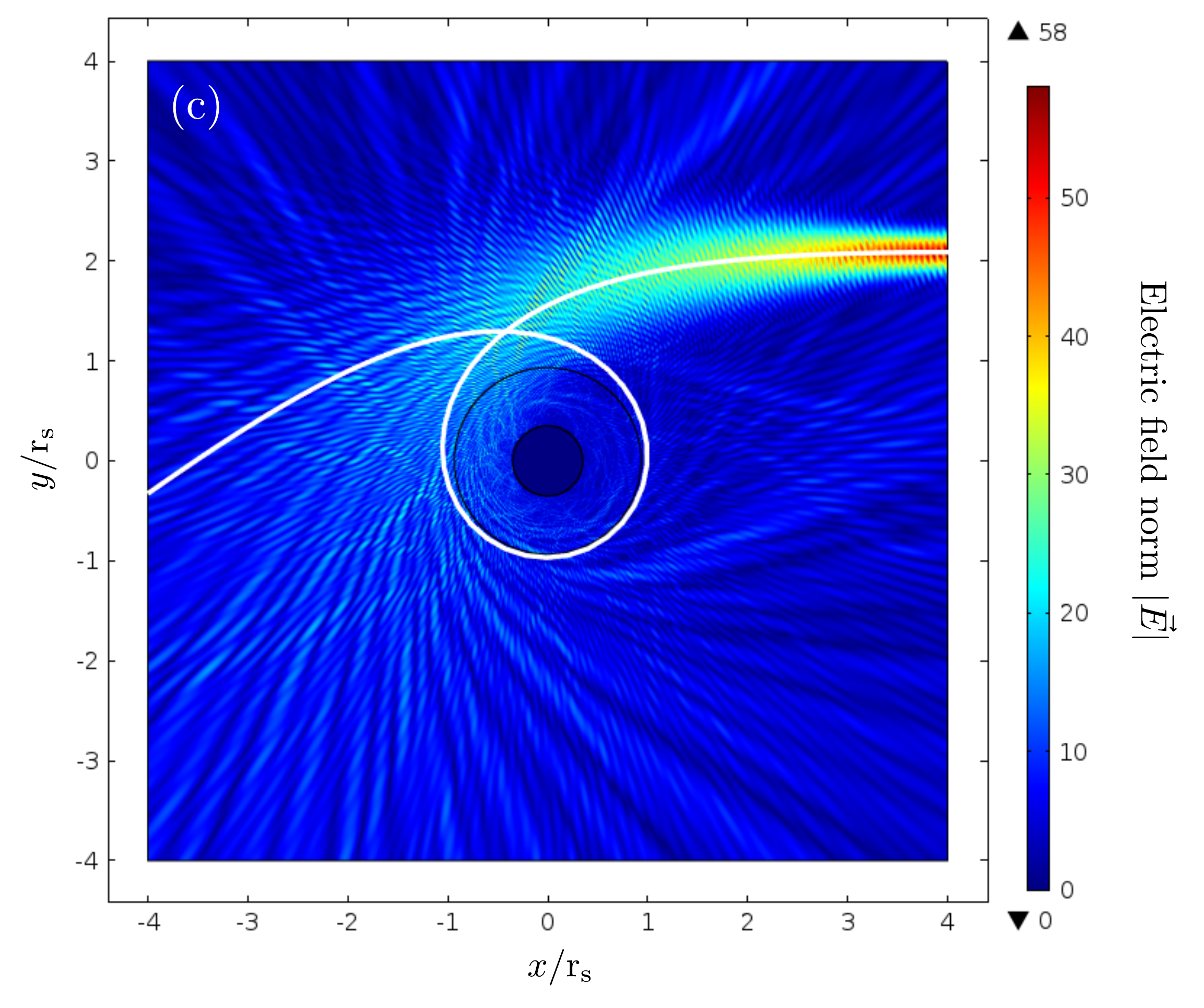}}
\subfigure{\includegraphics[width=0.85\textwidth]{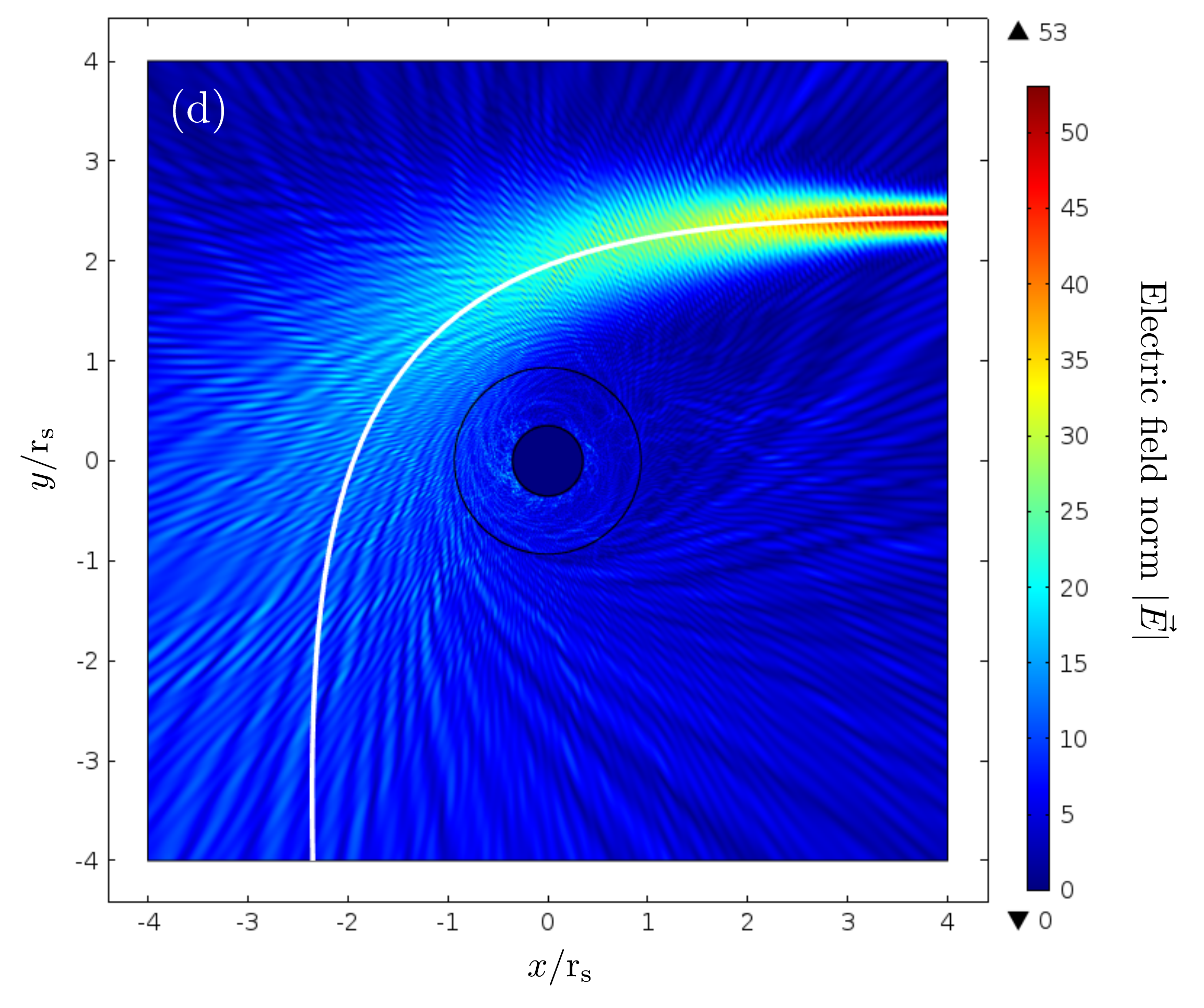}}
\caption{
TE Gaussian beam of wavelength $\lambda=0.15\rs$ compared with ray path
(superimposed in white) in metamaterial mimicking the Schwarzschild isotropic
spacetime for different impact parameters:
(a) $b=\rs$, (b) $b=2\rs$, (c) $b\approx b_{\rm c}$ (slightly above), and (d)
$b=3\rs$.
The event horizon $\rm{\rho_s}$ and the photon sphere of radius
$(2+\sqrt{3})\rs/4$ are depicted by black circles.}
\label{fig:iso}
\end{figure}

We solve Eq.\ \eqref{eq:te} in a rectangular 2D geometry of $(x,y)$ space for a
TE polarized wave injected from the right
(see Figs.\ \ref{fig:schw} - \ref{fig:iso}).
The medium parameters we use are given by Eq.\ \eqref{eq:perm-sch} for the
anisotropic spacetime and by Eq.\ \eqref{eq:n-iso} for the isotropic spacetime.
In both cases, we solve the equations in Cartesian coordinates.
The computational domain is surrounded by a perfectly matched layer that absorbs
the outward waves to ensure that there are no unwanted reflections, and the
simulations are done by means of a standard software solver.

It is known that the Schwarzschild black hole presents an event horizon at
$r=\rs$ (in isotropic coordinates, at $\rho = \rs/4 \equiv \rm{\rho_s}$)
where the gravity is so strong, that light cannot escape \cite{chandra}.
The metric has a coordinate singularity there,
which translates into singular medium parameters.
To avoid it, we set an effective horizon at $\rs+\delta$ with $\delta$ being
small positive number
\footnote{We set $\delta=0.05\,\rs$ (anisotropic case) and $\delta=0.11\,\rs$
(isotropic) in the calculations. 
We have checked that the variation of $\delta$, whenever it is small enough,
does not affect the global distribution of fields in the whole domain except
some tiny shell at the horizon.}
and impose an absorbing inner core for $r<\rs+\delta$
\cite{narimanov09, chen10}.

We use a Gaussian shape for the TE wave injected at the boundary, however, care
should be taken for the injection direction. If the medium is anisotropic, the
beam does not propagate in the direction of the wave normal, as we discussed in
Sec.\ \ref{TE-TM}. There will be an angle $\alpha$ between the wave velocity
and the ray velocity (see Fig.\ \ref{fig:angle}).
To ensure that the wave is always injected in a controlled way, in our case
along the $x$-axis, independently of the point of injection (that is, for
different impact parameters $b$), we impose:
$\vec{n}= - (\cos\alpha) \vec{\hat{x}} - (\sin\alpha) \vec{\hat{y}}$
for the normal unit vector $\vec{n}$ of the wave front.
The angle $\alpha$ is determined by Eq.\ \eqref{tan} applied to the point
of injection $(x_0,y_0)$.
The fields correspond to those depicted in Fig.\ \ref{fig:angle-te}, in
particular, the electric field vector is perpendicular to the plane of
simulation and the medium has a magnetic anisotropy.
Figure \ref{fig:angle-tm} corresponds to the case of TM wave injection into
an electrically anisotropic medium. This case, in principle, can also be
simulated. Because the injected wave is not a plane wave, the wavevector
$\vec{k}$ in Fig.\ \ref{fig:angle} should be replaced by the wave normal vector
$\vec{n}$ of the Gaussian beam.

The impact parameter $b$ is the key quantity which allows us to distinguish
between different types of ray trajectories \cite{chandra}.
Depending on its value, one may observe capture ($b < b_{\rm c}$)
or deflection ($b > b_{\rm c}$).
Its critical value $b_{\rm c}=3\sqrt{3}\,\rs/2$ determines the so-called
photonic sphere, which corresponds to an unstable circular orbit of radius
$3\,\rs/2$. To relate the value of $b$ with the initial coordinates
$(x_0,y_0)$, we use the following formula
\begin{equation}
b = y_0 \sqrt{ - \frac{\grr}{\gtt}}
\frac{ r_0 f}{\sqrt{ x_0^2 + f y_0^2 }}
\label{b-0}
\end{equation}
with the metric parameters taken at $r_0=\sqrt{x_0^2+y_0^2}$. Equation
\eqref{b-0} has been derived under the assumption $\dot{y}=0$ at the injection
point. It can be verified that at infinity ($r_0\to\infty$) $b$ approaches $y_0$
as it should according to the definition of the impact parameter.
In anisotropic spacetimes, the injection momentum
$\vec{p}= p_x \vec{\hat{x}} + p_y \vec{\hat{y}}$
is not collinear with the propagation direction.
Since we impose that initially the ray propagates along the $x$-axis, at the
injection point the momentum forms an angle given by
\begin{equation}
\frac{p_{y0}}{p_{x0}} = \frac{(1-f) x_0 y_0}{x_0^2+f y_0^2} = \tan\alpha.
\label{tan-p}
\end{equation}
It is exactly the same angle $\alpha$ between the wave normal and the
Poynting vector and between the $\vec{B}$ and $\vec{H}$ fields, all given by
\eqref{tan}. Here we clearly see how the spacetime anisotropy results in the
anisotropy of the medium.

In Fig.\ \ref{fig:schw} we show the results of numerical simulations for
Gaussian beams incident with different impact parameters $b$ for the case of a
Schwarzschild anisotropic spacetime.
The ray paths calculated from the Hamilton equations are superimposed in
white on the wave pattern (shown in color) simulated by solving the wave
equation.
We observe a good correspondence between the two approaches: the ray path
follows the center of the Gaussian beam even when it deflects.
Moreover, the ray trajectories are consistent with the behavior expected from
theory: we observe the ray capture for
$b < b_{\rm c}$ [Figs.\ \ref{fig:schw}(a),(b)], and its deflection for
$b > b_{\rm c}$ [Figs.\ \ref{fig:schw}(c),(d)]. The wave solutions present a
more complex behavior due to interference effects.
The Gaussian beam splits into a set of rays or ``subbeams'', one part of it is
captured due to having $b<b_{\rm c}$ and another part bends around the photonic
sphere of the black hole and interferes with the primary beam.
This effect is more pronounced in Figs.\ \ref{fig:schw}(b),(c) for the values
of $b$ which are not very far from the critical value $b_c\approx 2.598r_s$.
It is interesting to see that the subbeams are regularly spaced while bending
around the black hole and some of them exhibit smaller subwavelength ripples.
Similar behavior was observed in Ref.\ \cite{chen10} for a larger value of the
wavelength.

In Fig.\ \ref{fig:iso} we present the results for the isotropic case of the
Schwarzschild metric using the same set of the impact parameters $b$ as for the
anisotropic case of Fig.\ \ref{fig:schw}. Again, we observe a good agreement
between the wave propagation and the ray tracing.
The principal difference is that now the local velocity of light becomes
isotropic and it is determined by a scalar function -- the refractive index
$n(\rho)$.
As a consequence, the interference pattern we observe is substantially
different. One can see irregular ``jets'' leaving the system in radial
directions, similar to those obtained by Genov et al.\ for the isotropic
spacetimes \cite{genov09}. The small ripples are also clearly seen.
Since we simulate an isotropic medium, another distinctive feature we observe,
is that the wavefronts are perpendicular to the beam propagation, i.e. 
$\alpha=0$. 

On the other hand, if we compare the ray trajectories for the isotropic and the
anisotropic cases, they are rather similar: the same kinds of orbits (capture
or deflection) are observed corresponding to the same value of the impact
parameter $b$.
Notice that the parameter $b$ is a coordinate-independent quantity, since it is
defined as a ratio between the angular momentum and energy of the photon
measured at infinity.
Other physical properties, like the deflection angle between the incoming and
outgoing light rays measured by a distant observer at infinity should also be
equal independently of the coordinate system used.

\clearpage

\begin{figure}
\centering
\subfigure{\includegraphics[width=0.85\textwidth]{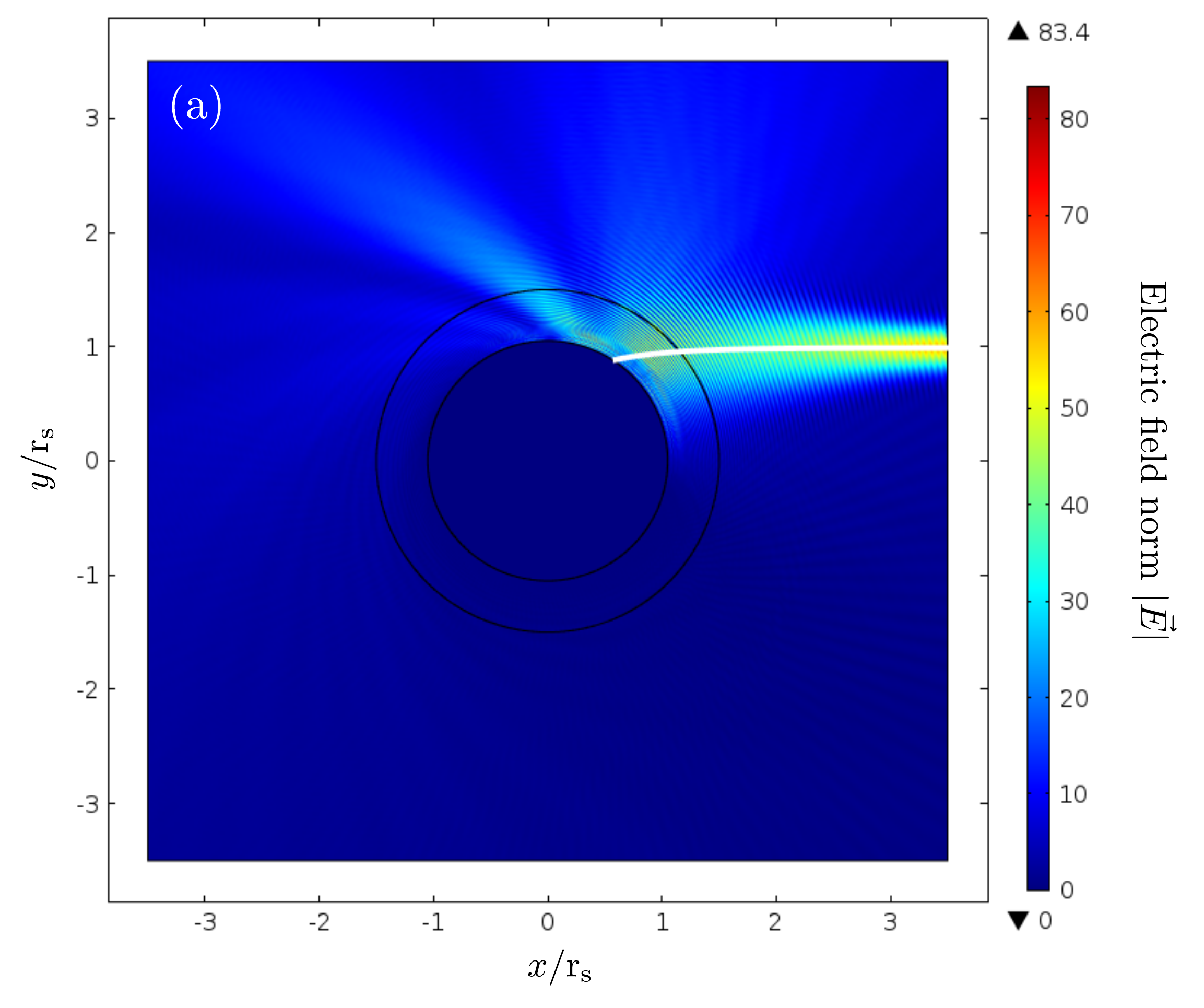}}
\subfigure{\includegraphics[width=0.85\textwidth]{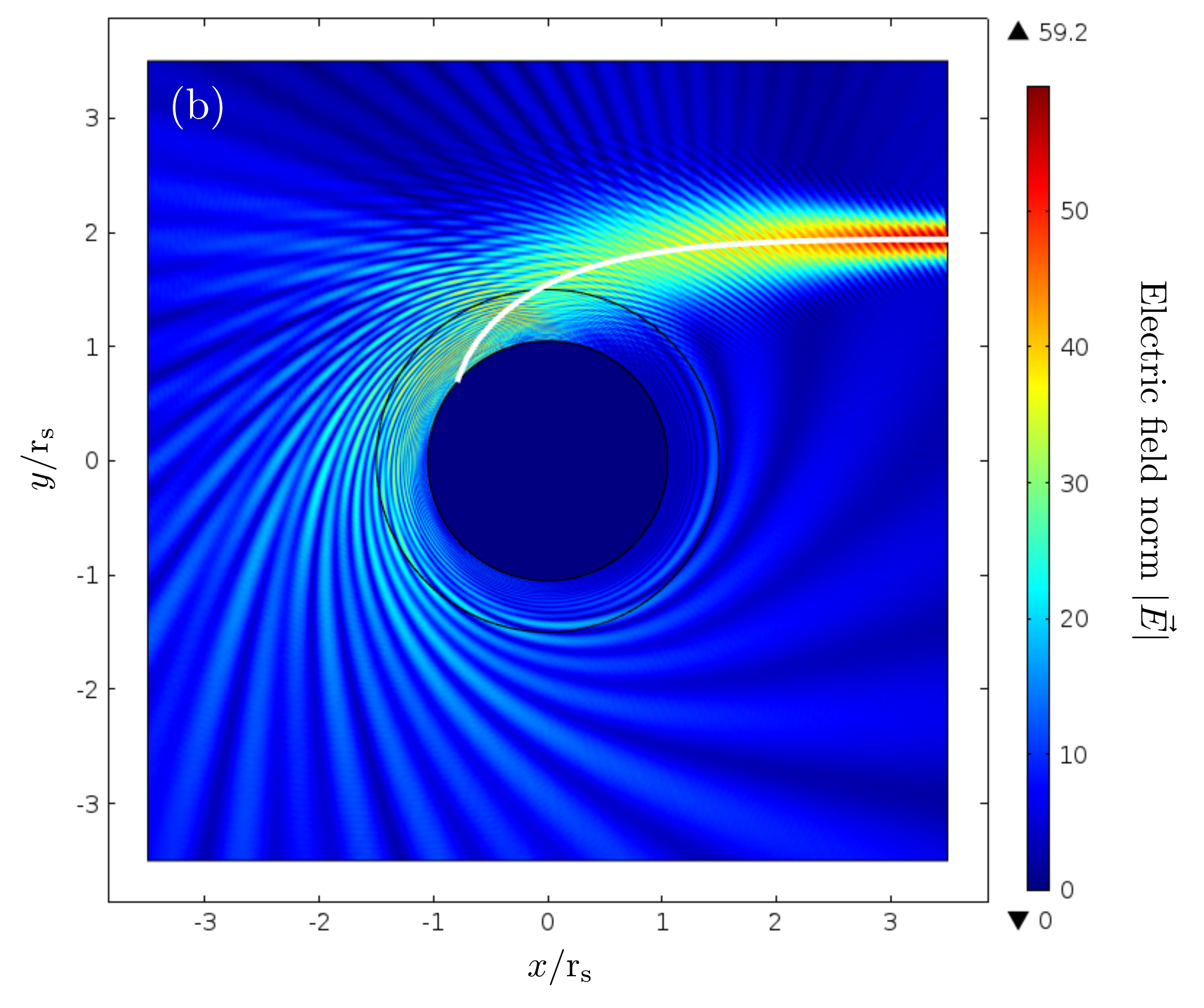}}
\end{figure}

\begin{figure}
 \centering
\subfigure{\includegraphics[width=0.85\textwidth]{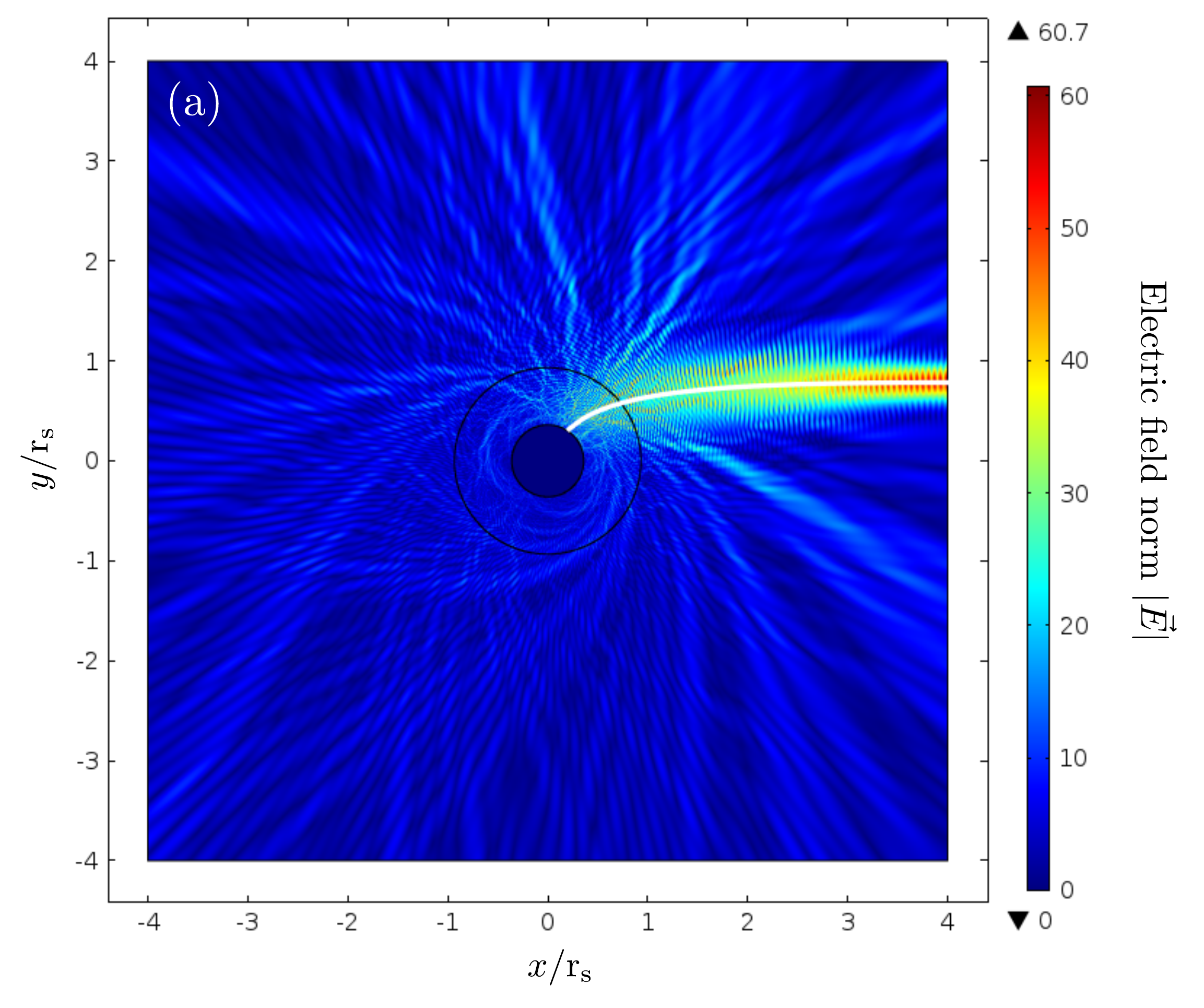}}
\subfigure{\includegraphics[width=0.85\textwidth]{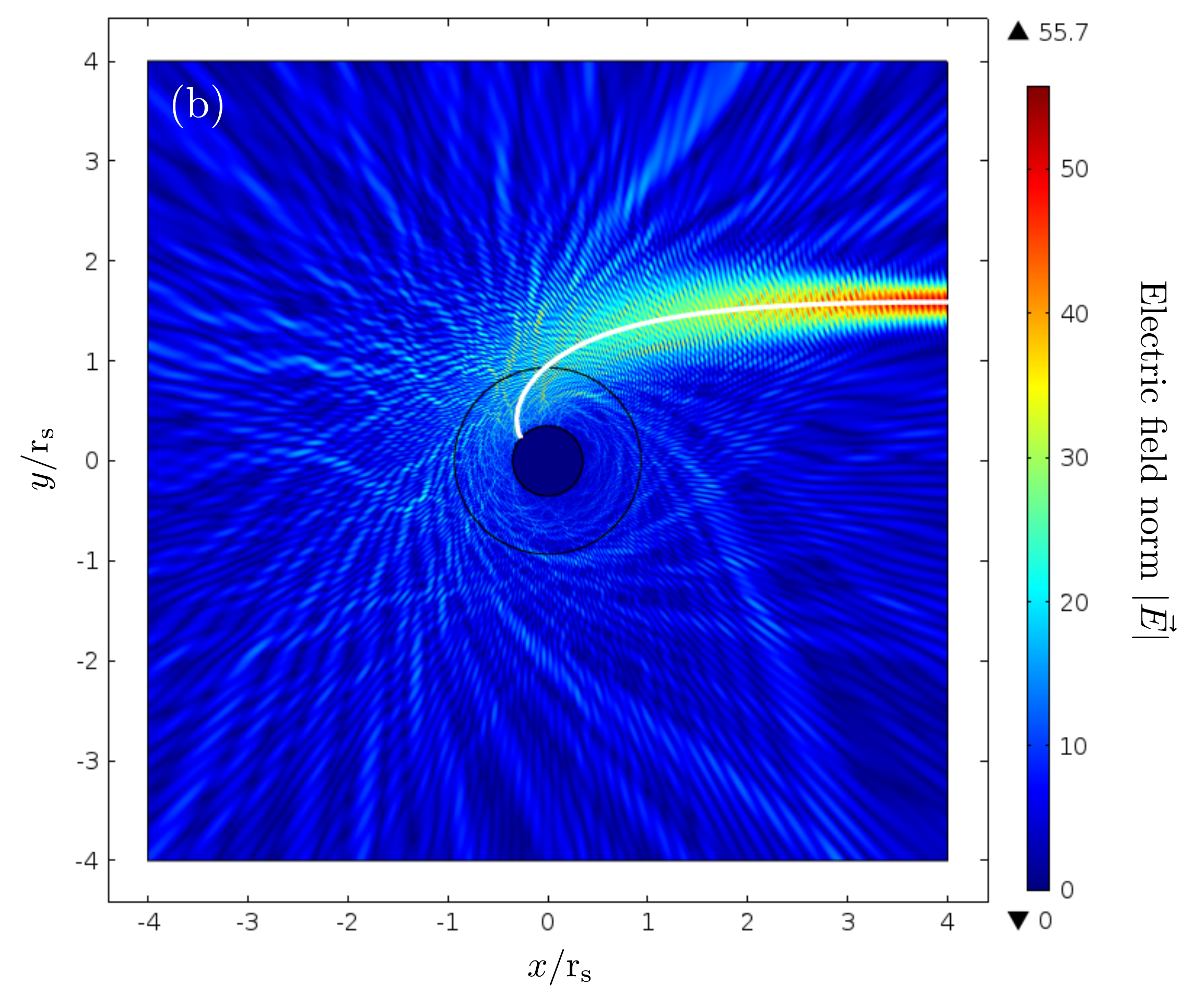}}
\end{figure}

\clearpage

\section{Concluding remarks}

In conclusion, we have derived the constitutive relations of an inhomogeneous
anisotropic medium which is formally equivalent to a class of static spacetime
metrics associated with a spherically symmetric cosmological black hole.
It is known that every spacetime metric can be written in different coordinate
systems and therefore, can be projected in many different ways into the
corresponding effective medium, each with specific material properties.
This procedure is at the basis of the transformation optics \cite{leonhardt0609,
leonhardt10}.
In particular, a static spherically symmetric spacetime written in arbitrary
(non-isotropic) coordinates can be equivalently described in isotropic
coordinates (conformally flat form) after performing the appropriate coordinate
transformation.

We have analyzed the medium properties and the propagation of light through the
effective media corresponding to these two cases of interest: anisotropic and
isotropic.
For the anisotropic spacetime, we found that only one kind of the medium
anisotropy would be essential: either magnetic or electric, for TE or TM
polarized waves respectively.
In both cases, light does not propagate in the direction of the wave normal.
There appears an angle between the wave velocity and the ray velocity,
specified by Eq.\ \eqref{tan}, which is related to the anisotropic factor of the
metric.
For some spacetimes, this angle can even be negative giving such interesting
phenomenon as negatively refracting medium \cite{lakhtakia05}.
However, this cannot be observed for the metric \eqref{eq:stat-spher-metric}.

It is interesting to note that for the kind of metric we consider, the time and
space metric coefficients correspond to different properties of the medium:
either electric or magnetic. For instance, for the TE wave, the time dilation
term determines the dielectric permittivity of the equivalent medium, while the
spatial metric components correspond to its magnetic permeability.
This phenomenon cannot be seen in the isotropic medium characterized by the
scalar refractive index.
We presented the results for the TE wave in the magnetically anisotropic media,
although the case of the TM wave in the electrically anisotropic media perhaps
would be easier to realize in the laboratory experiment. Both considerations
are theoretically equivalent.

By applying the eikonal approximation to the wave equation, we have also
obtained the expression for the optical Hamiltonian which was used to simulate
the ray paths. It coincides with the Hamiltonian obtained from general
relativity for null-geodesics.
It is clearly seen from our theoretical approach, that despite the fact that in
the geometrical optics limit different media may give the same ray paths, their 
full-wave description will differ.
Taking as an example the Schwarzschild spacetime metric, we obtained a very
good correspondence between the wave propagation and the ray trajectories for
the anisotropic and isotropic media.
Although the isotropic case is easier to implement in metamaterials, we believe
that due to a rapid progress in metamaterial technology \cite{monticone14}
(see also proposal \cite{mackay11}), anisotropic media will also be available to
model the cosmological phenomena in the laboratory and many physically
interesting phenomena could be observed.

\section*{Acknowledgements}

IFN acknowledges  financial  support  from  Universitat  de  Barcelona  under
the APIF scholarship.

\end{document}